\documentclass[final,authoryear,3p,times,twocolumn]{elsarticle}

\usepackage{journals}
\usepackage{graphicx}
\usepackage{epsfig}
\usepackage{color}
\usepackage{amsmath,amssymb,latexsym}

\journal{New Astronomy}

\begin{document}
\begin{frontmatter}
\title{A Natural Symmetrization for the Plummer Potential}

\author[NAOJ]{Takayuki R.Saitoh}
\ead{saitoh.takayuki@nao.ac.jp}
\author[NAOJ,CfCA,SOKENDAI]{Junichiro Makino}
\ead{makino@cfca.jp}
\address[NAOJ]{Division of Theoretical Astronomy, National Astronomical
Observatory of Japan, 2--21--1 Osawa, Mitaka-shi, Tokyo 181--8588.}
\address[CfCA]{ Center for Computational Astrophysics, National Astronomical
Observatory of Japan, 2--21--1 Osawa, Mitaka-shi, Tokyo 181--8588}
\address[SOKENDAI]{Department of Astronomical Science, School of Physical
Sciences, The Graduate University for Advanced Studies (SOKENDAI), 2--21--1
Osawa, Mitaka-shi, Tokyo 181--8588, Japan.}

\begin{abstract}
We propose a symmetrized form of the softened gravitational potential which is a
natural extension of the Plummer potential. The gravitational potential at the
position of particle $i$ $(x_i,y_i,z_i)$, induced by particle $j$ at
$(x_j,y_j,z_j)$, is given by:
\begin{equation*}
\phi_{ij} = -\frac{G m_j}{|{r_{ij}}^2+ {\epsilon_{i}}^2+{\epsilon_{j}}^2|^{1/2}}, 
    \label{eq:abs}
\end{equation*}
where $G$ is the gravitational constant, $m_j$ is the mass of particle $j$,
$r_{ij} = | {(x_i - x_j)^2 + (y_i - y_j)^2 + (z_i - z_j)^2} |^{1/2}$ and 
$\epsilon_i$ and $\epsilon_j$ are the gravitational softening lengths of
particles $i$ and $j$, respectively.  This form is formally an extension of the
Newtonian potential to five dimensions. The derivative of this equation in the
$x,y$, and $z$ directions correspond to the gravitational accelerations in these
directions and these are always symmetric between two particles.

When one applies this potential to a group of particles with different softening
lengths, as is the case with a tree code, an averaged gravitational softening
length for the group can be used.  We find that the most suitable averaged
softening length for a group of particles is $\langle {\epsilon_j}^2 \rangle =
\sum_j^N m_j {\epsilon_j}^2 / M$, where $M = \sum_j^N m_j$ and $N$ are the mass
and number of all particles in the group, respectively.  The leading error
related to the softening length is $\mathcal O \left (\sum_j {{\boldsymbol r}_j
d(\epsilon_j}^2)/{{r_{ij}}^{3}} \right )$, where ${\boldsymbol
r}_j$ is the distance between particle $j$ and the center of mass of the group and
${d(\epsilon_j}^2) = {\epsilon_j}^2 - \langle {\epsilon_j}^2
\rangle$.  
Using this averaged gravitational softening length with the tree method, one can
use a single tree to evaluate the gravitational forces for a system of particles
with a wide variety of gravitational softening lengths.  Consequently, this will
reduce the calculation cost of the gravitational force for such a system with
different softenings without the need for complicated forms of softening. We
present the result of simple numerical tests. We found that our modification of
the Plummer potential works well.
\end{abstract}

\end{frontmatter}

\section{Introduction} \label{sec:introduction}

\cite{Aarseth1963} introduced a Plummer potential \citep{Plummer1911} for 
gravitational $N$-body simulations in order to eliminate the divergence of the
potential that occurs when two particles undergo a close encounter.  With a
Plummer softening, the modified gravitational potential between two particles
$i$ and $j$ is given by 
\begin{equation}
{\phi}_{ij} = -\frac{G m_j}{|{{\boldsymbol r}_{ij}}^2 + 
{\epsilon}^2|^{1/2}}, \label{eq:plummer_pot}
\end{equation}
where $G$, ${\boldsymbol r}_{ij}$, $m_j$, and $\epsilon$ are the gravitational
constant, the distance between particles $i$ and $j$, the mass of particle $j$
and the gravitational softening length, respectively.  The increase in the
calculation cost due to this modification of the potential is quite small.
Thus, it is widely used in simulations of self-gravitating systems and the GRAPE
(GRAvity PipE) series also adopts this form of the gravitational potential
\citep{Sugimoto+1990, Ito+1991, Okumura+1993, Makino+1997, Kawai+2000,
Makino+2003, KawaiFukushige2006, Makino2005, Makino2007}.  Due to the wide
adoption of the Plummer potential, there have been several studies on the choice
of the optimal gravitational softening length \citep{Merritt1996,
Athanassoula+2000, Dehnen2001}.

In simulations of self-gravitating systems, it is necessary to deal with a wide
range in both space and time. For example, one needs to cover the range of
$\sim 100~{\rm Mpc}$ (for the tidal field) to $\sim 10~{\rm pc}$ or less (for
the structure of the interstellar medium) in galaxy formation simulations .
Usually, the outer region of the target galaxy is represented by massive
particles in order to reduce the calculation cost
\citep[e.g.,][]{NavarroBenz1991, KatzWhite1993}.  In simulations of the
coevolution of the galactic center and star cluster, it is necessary to use
different masses for different components, since required accuracies and spatial
resolutions are different \citep{Fujii+2007, Fujii+2010}.  Thus we need to use
different gravitational softening lengths for different types of particles.

If one uses the direct summation method to calculate the gravitational
force, it is possible to use an arbitrary symmetric form for the softening
length; for example, $[({\epsilon_i}^2+{\epsilon_j}^2)/2]^{1/2}$,
$(\epsilon_i+\epsilon_j)/2$, or $\max(\epsilon_i,\epsilon_j)$.  On the other
hand, when one uses the tree method \citep{Appel1985,BarnesHut1986}, one needs
to use separate trees for each of the different groups of particles with 
the same gravitational softening length \citep[e.g.][]{Springel+2001}, 
since otherwise there will be an error in the force calculation of the order
$\epsilon^2$. If one were to calculate the force from the center of mass of two
particles with widely different gravitational softening values, it is not
clear how one would take into account the gravitational softening length.
However, if one uses different trees for particles with different
gravitational softening lengths, the calculation cost increases by a large
factor.

\cite{HernquistKatz1989} proposed a different approach for softening.  They
assumed that each particle has a finite size and mass distribution given by a
function with a compact support, e.g. a spline function.  The advantage of this
gravity kernel is that it becomes a purely Keplarian potential for distances
larger than the particle size.  Thus, for particles separated by distances
larger than their sizes, one can use the center-of-mass approximation or the
multipole expansion of a purely Keplarian potential.  However, spline functions
require a few conditional branches and the evaluation of a polynomial.  These
operations lead to a non-negligible slowdown in the force and potential
calculations on modern CPUs/GPUs which have SIMD execution units.  As a result,
the calculation cost of a spline function is higher than that of the simple
Plummer softening.  Moreover, the tree approximation cannot be used for
particles within the softening length. Thus, if there are many particles within
the softening length, they can cause a large increase in the calculation cost.

In this paper, we propose a symmetrized gravitational potential between two
particles with different softenings, which is a natural extension of the Plummer
potential.  We derive the multipole expansion of a group of such particles with
this symmetrized potential. We then show that, in terms of accuracy, it is the
best to use the second moment of the softening lengths of the particles in order
to obtain the averaged softening length for the group.  The multipole moment up
to the second order has a simple form, which is easy and inexpensive to
calculate.

This paper is structured as follows: in section \ref{sec:sym}, we present a
symmetrized gravitational potential for a pair of particles with different
gravitational softening lengths and then introduce an averaged gravitational
softening length for a group of such particles.  The use of this potential with
the tree method is also discussed in the section.  Results of numerical tests
are given in section \ref{sec:tests}.  A summary and discussion are shown in
section \ref{sec:discussion}.

\section{Symmetrized Plummer Potential and its Applications to the Tree Code} \label{sec:sym}

\subsection{Symmetrized Pairwise Potential} \label{sec:sym_pot}
We modify the Plummer potential (Eq. \ref{eq:plummer_pot}) so that it is
symmetric between two particles even when their gravitational softening lengths
are different.  Here, we consider the following form:
\begin{equation}
{\phi}_{ij} = -\frac{G m_j}{|{{\boldsymbol r}_{ij}}^2
    +{\epsilon_i}^2+{\epsilon_j}^2|^{1/2}}. \label{eq:pot_sym}
\end{equation}
Formally, this form can be regarded as the $1/r$ potential in the five
dimensional space ($x,y,z,\epsilon_i,\epsilon_j$).  This form has been used
before to model the gravitational interaction between galaxy particles
with different size and mass \citep{White1976, AarsethFall1980}.  When we take
the gradients of Eq. \ref{eq:pot_sym} in the $x,y$ and $z$ directions, we obtain
the gravitational accelerations, {\it i.e.}
\begin{eqnarray}
a_{{ij},x} &=& -\frac{\partial \phi_{ij}}{\partial x}, \nonumber \\
&=& -\frac{G m_j(x_i-x_j)}{|{{\boldsymbol r}_{ij}}^2 
    +{\epsilon_i}^2+{\epsilon_j}^2|^{3/2}}. \label{eq:acc_sym}
\end{eqnarray}
This gravitational acceleration satisfies Newton's third law.

\subsection{Multipole Expansion of the Potential of a Group of Particles with
the Symmetrized Plummer Potential
} \label{sec:multipole}

In this section, we derive the multipole expansion of the gravitational
potential for a group of particles with the symmetrized Plummer potential by
introducing an averaged softening for the group of particles.  Using Eq.
\ref{eq:pot_sym}, we obtain the most suitable form of the averaged softening
which minimizes the potential/force error. This form can easily be used with a
tree code \citep{BarnesHut1986}.

We consider the gravitational potential at ${\boldsymbol r}_i$ due to a group of
particles $j = 1, \dots, N$ with the total mass $M$. The three-dimensional
position of particle $j$ is expressed as ${\boldsymbol {r}}_j$. Here we choose
the center of mass of particles $j$ as the coordinate center.  By
taking the summation over $j$, the potential at ${\boldsymbol r}_i$ is 
\begin{equation}
{\phi}_i = -\sum_j^N \frac{G m_j}{|{{\boldsymbol r}_{ij}}^2
    +{\epsilon_i}^2+{\epsilon_j}^2|^{1/2}}, \label{eq:pot_sym_sum}
\end{equation}
where ${\boldsymbol r}_{ij} = {\boldsymbol r}_i - {\boldsymbol {r}}_j$.

We consider the following form as the monopole potential of the group:
\begin{equation}
{\phi'}_i = -\frac{G M}{|{{\boldsymbol r}_i}^2 +{\epsilon_i}^2+ {\mathcal{E}_j}^2 |^{1/2}},
    \label{eq:pot_sym_sum_averaged}
\end{equation}
where ${\mathcal{E}_j}^2$ is an arbitrary form of the averaged softening length for
the group of particles $j$. As shown below, in terms of accuracy, we can
obtain the most suitable form of the averaged softening length for the group of
particles.

By using ${\mathcal{E}_j}^2$, we can
rewrite Eq. \ref{eq:pot_sym_sum} as follows:
\begin{equation}
{\phi}_i = -\sum_j^N \frac{G m_j}{|({{\boldsymbol r}_{i}-{\boldsymbol {r}}_{j}})^2
    +{\epsilon_i}^2+ {\mathcal{E}_j}^2 +d({\epsilon_j}^2)|^{1/2}},
    \label{eq:pot_sym_sum_diff}
\end{equation} 
where $d({\epsilon_j}^2) = {\epsilon_j}^2-{\mathcal{E}_j}^2$.
The Taylor expansion of Eq. \ref{eq:pot_sym_sum_diff} up to the second order of
${\boldsymbol {r}}_j$ and $d({\epsilon_j}^2)$ is 
\begin{eqnarray}
{\phi}_i &=& 
-\sum_j^N \frac{G m_j}{R} \Biggl \{ 1
        + \frac{({\boldsymbol r}_i\cdot{\boldsymbol {r}}_{j})}{{R}^2}
+ \frac{d({\epsilon_j}^2)}{{R}^2}
        \nonumber \\
&&+ \frac{3 ({\boldsymbol r}_i\cdot{\boldsymbol {r}}_{j})^2 
    - |{\boldsymbol r}_i|^2 |{\boldsymbol {r}}_{j}|^2}{{R}^4}
+ \frac{3 ({\boldsymbol r}_i\cdot{\boldsymbol {r}}_{j})
    d({\epsilon_j}^2)}{{R}^4} \nonumber\\
&&+\frac{3 \left[ d({\epsilon_j}^2) \right]^2}{4 {R}^4} 
+ \mathcal O \left( \left[ \frac{{\boldsymbol
        {r}}_{j}}{R}+\frac{d({\epsilon_j}^2)}{R^2} \right]^3 \right) \Biggl \}
\label{eq:pot_sym_sum_exp} 
\end{eqnarray} 
where 
\begin{equation}
R = \left( {{\boldsymbol r}_i}^2+{\epsilon_i}^2+ {\mathcal{E}_j}^2 \right)^{1/2}.
\label{eq:Plummer_distance}
\end{equation}
The zeroth term of the Taylor expansion is ${\phi'}_i$ itself (Eq.
\ref{eq:pot_sym_sum_averaged}).  The term of the order of $({\boldsymbol
r}_i\cdot{\boldsymbol {r}}_{j})$ vanishes by definition.  Since
$d({\epsilon_j}^2) = {\epsilon_j}^2-{\mathcal{E}_j}^2$, the first order term of
$d({\epsilon_j}^2)$ vanishes if we adopt the second moment of $\epsilon_j$ as
the averaged softening length:
\begin{equation}
{\mathcal{E}_j}^2 = \langle \epsilon_j^2 \rangle \equiv \frac{\sum_j^N m_j {\epsilon_j}^2 }{M}.
\label{eq:averaged_e}
\end{equation}
Hence, choosing the second moment of $\epsilon_j$ to be the averaged softening
length is the most favourable option; using other forms such as
$(\epsilon_i+\epsilon_j)/2$, $\max(\epsilon_i,\epsilon_j)$ and
$\min(\epsilon_i,\epsilon_j)$ result in a second-order error term.

For the convergence of the multipole expansion, it is necessary to take into
account not only the ordinary three-dimensional distance but also the
distribution of the softenings, $\epsilon_j$.  We will discuss this point in the
next subsection.

\subsection{The opening criterion} \label{sec:tree}

In the tree method, the force and potential of a group of particles are
approximated by that of the corresponding tree node.  The opening criterion
\citep{BarnesHut1986},
\begin{equation}
\theta > \frac{w}{d}, \label{eq:tolerant}
\end{equation}
is widely used to determine whether the multipole approximation can be used or
not. Here, $\theta$ is called the opening angle and controls the force/potential
accuracies, $w$ is the size of the tree node and $d$ is the distance between
the barycenter of the tree node and the position where we want to calculate the
force and potential. Since the multipole expansion of the $1/r$ potential around
$r+dr$ is a series of $dr/r$ where $dr \sim w$ and $r \sim d$, Eq.
\ref{eq:tolerant} corresponds to the convergence condition for the multipole
expansion of the $1/r$ potential. For the Plummer potential, the Plummer
distance, $(r^2 + {\epsilon_i}^2)^{1/2}$, is adopted instead of $r$. Typically,
a value of $\theta = 0.5 \sim 0.75$ is used.  Note that, unless $\theta \le
1/\sqrt{3}$, the Barnes \& Hut's opening criterion causes unbound errors in
force calculation, resulting in ``detonating galaxies''
\citep{SalmonWarren1994}.  They studied other opening criteria so that one can
avoid these large errors. Here, we use Eq. \ref{eq:tolerant} with sufficiently
small $\theta$. Hence our results, which will appear in section
\ref{sec:EnergyError}, are free from this problem.

As shown in section \ref{sec:multipole}, the potential induced by a group of
particles with the symmetrized form of the potential is expanded by two
parameters, $ {\boldsymbol {r}}_j$ and ${d\epsilon_j}^2$. Thus, the following
two constraints together should satisfy the conditions for convergence:
\begin{equation}
\eta >  \frac{| {\boldsymbol {r}}_j |}{R},
\end{equation}
and 
\begin{equation}
\eta' >  \frac{d({\epsilon_j}^2)}{R^2}.
\end{equation}
Here, $\eta$ and $\eta'$ are the tolerance parameters for the multipole expansion
of the symmetrized Plummer potential.  By replacing ${\boldsymbol {r}}_j$
with $w$ and ${d\epsilon_j}^2$ with ${\epsilon_{\rm {max}}}^2-{\epsilon_{\rm
{min}}}^2$, where $\epsilon_{\rm {max}}$ and $\epsilon_{\rm {min}}$ are the
maximum and minimum softening lengths in a tree node, respectively, we
obtain 
\begin{equation}
\eta >  \frac{w}{R}, \label{eq:tolerant_sym_r}
\end{equation}
and 
\begin{equation}
\eta' >  \frac{{\epsilon_{\rm {max}}}^2-{\epsilon_{\rm {min}}}^2}{R^2}, 
    \label{eq:tolerant_sym_eps}
\end{equation}
for the convergence conditions of the multipole expansion.

Note that ${\epsilon_{\rm max}}^2 - {\epsilon_{\rm min}}^2$ corresponds to the size of the
``box'' in the fifth dimension. In our current implementation, the tree
structure is still an oct-tree in three dimensions. Thus, if a node contains two
particles which have widely different gravitational softening lengths, its ``size'' can
be dominated by the softening if its physical size is small. When we calculate
the force from this node to a nearby particle with a distance $\leq [({\epsilon_{\rm
max}}^2 - {\epsilon_{\rm min}}^2)/\eta']^{1/2}$, this node is always opened up even if the
physical size of the node in three dimensions is much smaller than the distance.
This behavior could cause an unnecessary increase in the calculation cost. In
principle, we can easily solve this problem by making a tree structure in 
four-dimensional space, but so far we have not used such treatment.

In the tree-with-GRAPE method \citep{Makino1991TreeWithGRAPE}, a similar opening
criterion, a set of Eqs \ref{eq:tolerant_sym_r} and \ref{eq:tolerant_sym_eps},
can be used with two modifications based on an argument for the convergence
condition of the Taylor expansion. The modifications are as follows: First, we
employ the minimum node-to-node distance in the ordinary three dimensions as the
separation distance, ${\boldsymbol r}_i$, in Eq. \ref{eq:Plummer_distance}.
Then, we use the minimum gravitational softening length in the particles which
share an interaction list, instead of $\epsilon_i$ in Eq.
\ref{eq:Plummer_distance}.  These modifications give a sufficient accuracy when
used with the tree-with-GRAPE method.

\subsection{Calculation of averaged softening} \label{sec:misc}

The averaged gravitational softening length for each tree node can be calculated
in the tree building phase with a recursive operation that is the same as that
for the multipole expansion. The averaged gravitational softening length for a
parent node, $\langle \epsilon_p^2 \rangle$, is 
\begin{equation}
\langle {{\epsilon}_{\rm p}}^2 \rangle
= \frac{\sum_k^{N_{\rm c}} M_{{\rm c},k}{{\langle {{\epsilon}_{{\rm c},k}}^2 \rangle }}}{M_{\rm p}},
\end{equation}
where $N_{\rm c}$ is the number of child nodes, $M_{\rm c,k}$ and $\langle
{{\epsilon}_{{\rm c},k}}^2 \rangle$ are the mass and the averaged gravitational
softening length of the $k$-th child node, respectively, and $M_{\rm p}$ is the
mass of the parent node.

\subsection{Higher order expansion and application to FMM} \label{sec:FMM}

In this section, we discuss the application of the symmetric softening to the
fast multipole method (FMM) \citep{GreengardRokhlin1987, Zhao1987, Dehnen2000}.
Unlike the tree method, in FMM, multipole expansions are first translated to
local expansion, which is then evaluated at the positions of multiple particles.
Thus, the calculation order in FMM is proportional to $N$.

The calculation cost in FMM depends on the expansion order of the multipole
moments in the local/external particles.  In the standard FMM
\citep{GreengardRokhlin1987}, spherical harmonics is used to express expansion
and thus the calculation cost is $\mathcal O(p^4 N)$, where $p$ is the order of
the multipole expansion.  \cite{ElliottBoard1996} successfully reduced the
calculation cost in FMM from $\mathcal O(p^4 N)$ to $\mathcal O(p (\log p)^{1/2}
N)$ by using a fast fourier transformation and optimal choice of the tree level.
\citet{Dehnen2000} proposed a Cartesian FMM, however, its calculation cost is
estimated as $O(p^6 N)$.  Again this cost can be reduced to $\mathcal O(p^{3/2}
(\log p)^{1/2} N)$ by following \cite{ElliottBoard1996}.

With the $1/r$ potential in the five dimensions presented here, one cannot use a
spherical harmonic function for the expansion.  Thus, it is necessary to use a
straightforward expansion such as what is used in the Cartesian FMM
\citep{Dehnen2000} to implement our symmetrized Plummer potential.  In the
multipole expansions, because it is necessary to expand the fourth or fifth
dimension for the local and external field, the calculation cost is $\mathcal
O(p^8 N)$ and the reduced calculation cost is $\mathcal O(p^2 (\log
p)^{1/2}~N)$.  When the multipole expansion is applied, the softening lengths of
the particles are usually smaller than the size of the node and therefore one
usually needs to take into account only the first cross term in Eq.
\ref{eq:pot_sym_sum_exp}.  In this case, the calculation cost is similar to that
of the Cartesian FMM.  Moreover, gravitational softening is used in
simulations where the required force accuracy is not very high.  Therefore the
multipole expansion can be truncated in relatively low orders.  Thus, the
increase in the calculation cost is not so large if one incorporates the
symmetrized Plummer potential into FMM.

\section{Tests} \label{sec:tests}

\subsection{Error in Acceleration} \label{sec:AccError}

Here, we present the measured acceleration error of the force calculated using a
tree and our symmetrized form of the Plummer potential.  We take two groups of
$5\times10^4$ particles, each with a different mass and softening length. The
particles are randomly distributed within a uniform sphere of radius $L = 1$.
Table \ref{tab:runs} shows the different combinations of masses and
gravitational softening lengths used for this test.  The accuracy of the
accelerations depends strongly on the distribution of particles
\citep{BarnesHut1989}. However, for simplicity, we consider only a uniform
sphere here.  In this test, we adopt the center-of-mass approximation for the
force calculation and we use the set of Eqs \ref{eq:tolerant_sym_r} and
\ref{eq:tolerant_sym_eps} for the opening criterion.  For this criterion, the
tolerance is always kept the same and is represented by $\theta$ instead of
$\eta$ and $\eta'$ for simplicity.

\begin{table*}
\begin{center}
\caption{The summary of the masses and gravitational softening lengths of the two
types of particles used in our test simulations. $m_1$ and $\epsilon_1$ refer to
the mass and gravitational softening length for the less massive particles,
whereas $m_2$ and $\epsilon_2$ are those of the massive particles. The number of
particles in each group is $5\times10^4$.
}\label{tab:runs}
\begin{tabular}{ccccc}
\hline
\hline
Mass ratio & $m_1$ & $\epsilon_1$ & $m_2$ & $\epsilon_2$ \\
\hline
$1:1 $ & $1.00 \times 10^{-5}$ & $6.79 \times 10^{-3}$ & $1.00 \times 10^{-5}$ & $6.79 \times 10^{-3}$ \\ 
$1:8 $ & $2.22 \times 10^{-6}$ & $4.11 \times 10^{-3}$ & $1.78 \times 10^{-5}$ & $8.22 \times 10^{-3}$ \\
$1:64$ & $3.08 \times 10^{-7}$ & $2.13 \times 10^{-3}$ & $1.97 \times 10^{-5}$ & $8.51 \times 10^{-3}$ \\
\hline
\end{tabular}\\
\end{center}
\end{table*}

We measure the mean acceleration error as a function of the opening angle for the
three different mass ratios: 1:1, 1:8 and 1:64 (see Table \ref{tab:runs}).  The
mean acceleration error for each case is given by the following equation:
\begin{equation}
\langle \Delta a (\theta) \rangle = \frac{1}{N} \sum^N_i \frac{\left | {\boldsymbol
    a}_i(\theta) - {\boldsymbol a}_{i,0} \right |}{\left |{\boldsymbol a}_{i,0} \right |},
\end{equation}
where ${\boldsymbol a}_i (\theta)$ is the acceleration vector of particle $i$
estimated with tolerance $\theta$ and ${\boldsymbol a}_{i,0}$ is that evaluated
for $\theta = 0$. We also measure the mean acceleration error which excludes the
error due to $d(\epsilon^2)$.  To do this, we build two trees, where each tree
consists of particles with the same mass and softening. We then calculate the
accelerations of the particles by adding the forces from the two trees so that
we obtain the mean acceleration error without the contribution of 
$d(\epsilon^2)$.

\begin{figure}[htbp]
\begin{center}
\includegraphics[width=0.46 \textwidth]{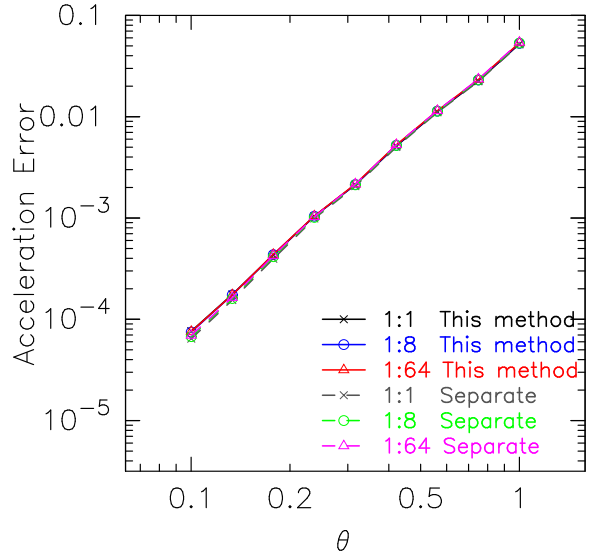}
\caption{Mean acceleration error as a function of the tolerance parameter
$\theta$.  Curves with crosses, circles and triangles are the results for
$m_2/m_1 = 1, 8$ and $64$, respectively.  The solid and dashed curves are the
results obtained by the tree which uses our averaged softening length (``This
method''), whilst the dashed curves are those obtained from using two separate
trees, where each tree consists of each particle group (``Separate'').
\label{fig:Acc_Errors}
}
\end{center}
\end{figure}

The mean acceleration error as a function of $\theta$, for each of these
methods, is shown in Figure \ref{fig:Acc_Errors}.  We found that there is no
clear difference in the mean acceleration errors between different methods.  We
also found that the mean acceleration error is approximately proportional to
$\theta^3$. This is in good agreement with  previous results obtained by the tree
code with the usual Plummer potential \citep{Hernquist1987, BarnesHut1989,
Makino1990}.  This means that our adopted opening criteria work correctly.

\subsection{Calculation Cost}

\begin{figure}[htbp]
\begin{center}
\includegraphics[width=0.46 \textwidth]{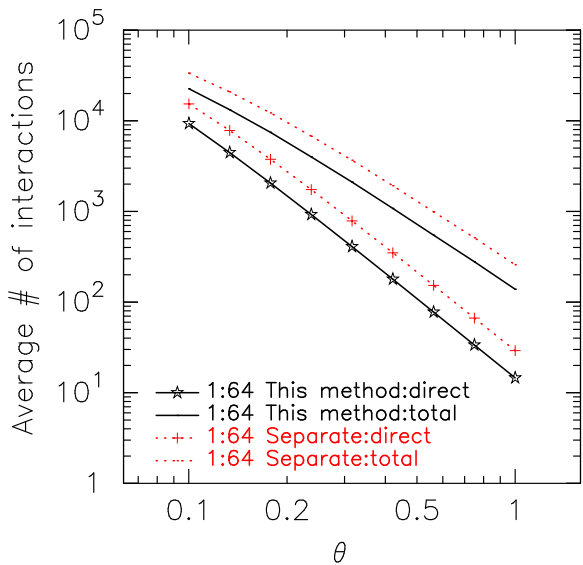}
\caption{Number of the average interactions as a function of $\theta$ with $L = 1$
and $m_2/m_1 = 64$.  Curves with symbols are the number of the mean
particle-particle interactions, whereas curves without symbols are the sum of
the number of the mean particle-particle and mean particle-node interactions.
\label{fig:InteractionList_L10}
}
\end{center}
\end{figure}

To evaluate the calculation cost, we measured the average number of
particle-particle interactions and the average number of total interactions.
Figure \ref{fig:InteractionList_L10} shows the these numbers as a function of
$\theta$.  We used the same distribution of particles as we used in section
\ref{sec:AccError}.  The dotted curves are the calculation cost for the case of two
separate trees for particles with different softenings. The calculation cost of our
method is about one half of that of the calculation with two separate trees.
The average number of the interactions is roughly proportional to $\theta^{-3}$
for both methods.  This dependence is again close to those in previous works
\citep{Makino1990}.

\subsection{Energy Error in Time Integration} \label{sec:EnergyError}

Here, we present the result of test calculations of a cold collapse.  We create a
uniform sphere with the radius $L$ and mass $M$ using a random distribution of 
$1024$ particles, where the system of units we use is $L=M=G=1$. 
The sphere consists of two groups of $512$ particles. The mass and softening
length of particles in these two groups are varied and the values used are given
in Table \ref{tab:runs2}.  The initial virial ratio of the system was set to
$0.1$ and the initial particle velocities were drawn from a Gaussian
distribution.

\begin{table*}
\begin{center}
\caption{The same as table \ref{tab:runs} but for the test calculations of cold
collapse.
}\label{tab:runs2}
\begin{tabular}{ccccc}
\hline
\hline
Mass ratio & $m_1$ & $\epsilon_1$ & $m_2$ & $\epsilon_2$ \\
\hline
$1:1$  & $9.77 \times 10^{-4}$ & $3.13 \times 10^{-2}$ & $9.77 \times 10^{-4}$ & $3.13 \times 10^{-2}$ \\
$1:8$  & $2.17 \times 10^{-4}$ & $1.89 \times 10^{-2}$ & $1.74 \times 10^{-3}$ & $3.79 \times 10^{-2}$ \\
$1:64$ & $3.00 \times 10^{-5}$ & $9.79 \times 10^{-3}$ & $1.92 \times 10^{-3}$ & $3.92 \times 10^{-2}$ \\
\hline
\end{tabular}\\
\end{center}
\end{table*}

The numerical simulations were performed using {\tt ASURA} (Saitoh, in
preparation). {\tt ASURA} adopts the tree-with-GRAPE method to calculate the
gravitational force and potential \citep{Makino1991TreeWithGRAPE}. For this test,
{\tt ASURA} only uses the center-of-mass of the tree nodes to calculate the
force, with their averaged gravitational softening lengths determined according
to Eq. \ref{eq:averaged_e} and the tolerance is set to $0.5$.  The maximum
number of particles which shares the same interaction list \citep{Barnes1990,
Makino1991TreeWithGRAPE} was set to $n_{\rm g} = 8$ and $n_{\rm g} = 128$.
Time-integration scheme used is an ordinary leap-frog method.

\begin{figure}[htbp]
\begin{center}
\includegraphics[width=0.46 \textwidth]{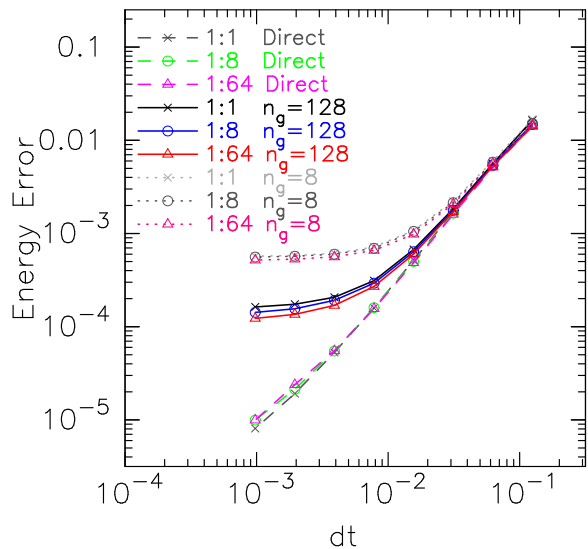}
\caption{Dependence of the relative energy error (at time $t = 1$) on time-step
size. Crosses, circles and triangles are the results for $m_2/m_1 =1, 8$ and
$64$, respectively.  Solid, dotted, and dashed curves are the results for
$n_{\rm g}=128$, $n_{\rm g}=8$, and direct calculation, respectively.
\label{fig:Errors}
}
\end{center}
\end{figure}

Figure \ref{fig:Errors} shows the relative error in the total energy of the
system, $\Delta E$, at time $t = 1$ (which is approximately the time of maximum
collapse) as a function of the size of the time-step.  Here $\Delta E$ is
defined as 
\begin{equation}
\Delta E = \left| \frac{E(t=0) - E(t=1)}{E(t=0)} \right|,
\end{equation}
where $E(t)$ is the total energy of the system at $t$.

From figure \ref{fig:Errors}, we can see that the mass ratio between groups of
particles does not affect the relative error.  For small time-steps ($\le
2^{-6}$), the relative error of the tree method is larger than that of the
direct method. This is because the error in the calculated force from using a
tree dominates the integration error.  The contribution of the tree
approximation to the relative error is smaller for large $n_{\rm g}$, since a
larger number of nearby particles is included in the direct calculation
\citep{Barnes1990}.

\section{Summary and Discussion}\label{sec:discussion}

In this paper, we introduced a natural symmetrization of the Plummer potential.
This symmetrization is quite simple and hence, easy for anyone to implement into
their own code, providing their code adopts the traditional Plummer potential
for gravity.

By applying this potential to a group of particles, we derived an averaged 
gravitational softening length for the group.  Consequently, we can use this
potential in the tree method.  The multipole expansion of the symmetrized
Plummer potential allows us to use a single tree to calculate forces and
potentials for a system of particles that have different softening lengths.
Thus the calculation cost of our method is less than that of previous
implementations.

Since the modification is quite simple, the modified Plummer potential can be
easily implemented in modern GRAPEs ({\it i.e.}, GRAPE-7 {\footnote {GRAPE-7
employed field programmable gate array (FPGA) and constructs the gravity
pipelines on runtime \citep{KawaiFukushige2006}.}} and GRAPE-DR {\footnote
{GRAPE-DR uses the original programmable SIMD processors \citep{Makino2005,
Makino+2007}.}}) as they can modify the force law via software libraries.
However, for older GRAPEs, the gravity pipelines are implemented on the hardware
layer and so it is impossible to use this modified potential.  The latest
version of the GRAPE-DR library supports this new symmetrized Plummer potential.

Pure software libraries, {\it i.e.} those that run on typical CPUs without
special hardware, can easily include our calculation model.  Phantom-GRAPE
(Nitadori et al, in preparation) is an optimized library that utilizes the SIMD
instructions of X86 CPUs and has application interfaces comparable with GRAPE-5
\citep{Kawai+2000}. For this software, we can easily modify the gravity kernel
in the library. We used this modified Phantom-GRAPE for the tests described in
Section \ref{sec:tests}.

The implementation of our modified Plummer potential on GPGPUs
\citep{Nakasato2009, Hamada+2009} is also straightforward.

\section*{Acknowledgements}
The authors thank an anonymous referee for his/her insightful comments which
helped us to improve our manuscript. The authors also thank Simon White and Piet
Hut for helpful discussion and William Robert Priestley for careful reading of
the manuscript.  Some of the numerical tests were carried out on Cray XT4 and
GRAPE system at the Center for Computational Astrophysics at the National
Astronomical Observatory of Japan.  This project is supported by Grant-in-Aid
for Scientific Researches (17340059 \& 21244020). TRS is financially supported
by a Research Fellowship from the Japan Society for the Promotion of Science for
Young Scientists.


\end{document}